\documentclass[10pt]{article}
\usepackage{latexsym}
\usepackage[dvips]{graphicx,color}
\newcommand{\be}{\begin{equation}}
\newcommand{\ee}{\end{equation}}
\def\n{\noindent}
\catcode `\@=11
\catcode `\@=12
\begin{document}
\begin{center}
\large{\bf {Plane Symmetric Inhomogeneous Cosmological Models with a Perfect Fluid in General 
Relativity}} \\
\vspace{10mm}
\normalsize{Anirudh Pradhan \footnote{To whom correspondence should be addressed}, 
Purnima Pandey$^2$ and Sunil Kumar Singh$^3$ } \\
\vspace{5mm}
\normalsize{$^{1,2}$ Department of Mathematics, Hindu Post-graduate College, 
Zamania-232 331, Ghazipur, India.} \\
\normalsize{$^1$E-mail : pradhan@iucaa.ernet.in} \\
\normalsize{$^2$E-mail : purnima\_pandey2001@yahoo.com}\\
\vspace{5mm}
\normalsize{$^3$ Department of Physics, S. D. J. Post-graduate College, Chandeswar-276 128,
Azamgarh, India.} 
\end{center}
\vspace{10mm}
\begin{abstract}
In this paper we investigate a class of solutions of Einstein equations for the plane- 
symmetric perfect fluid case with shear and vanishing acceleration. If these 
solutions have shear, they must necessarily be non-static. We examine the 
integrable cases of the field equations systematically. Among the cases with 
shear we find three classes of solutions.   
\end{abstract}
\n KEY WORDS : {Exact solutions, plane symmetric models, inhomogeneous universe} \\
\n PACS No. : 04.20.-q
\section{Introduction}

The standard Friedmann-Robertson-Walker (FRW) cosmological model prescribes a 
homogeneous and an isotropic distribution for its matter in the description of 
the present state of the universe. At the present state of evolution, the universe 
is spherically symmetric and the matter distribution in it is on the whole isotropic 
and homogeneous. But in its early stages of evolution, it could have not had such a 
smoothed out picture. Close to the big bang singularity, neither the assumption of 
spherically symmetric nor of isotropy can be strictly valid. From this point of view 
many authors consider plane symmetry, which is less restrictive than spherical 
symmetry and provides an avenue to study early days inhomogeneities. In recent years, 
cosmological models exhibiting plane symmetry have been studied by various authors 
(Rendall, 1995; Taruya and Nambu, 1996; Zhuravlev {\it et al.}, 1997; da Silva and Wang, 
1998; Ori, 1998; Anguige, 2000; Chervon and Shabalkin, 2000; Nouri-Zonoz and Tavanfar, 
2001; Yazadjiev, 2003; Pradhan {\it et al.}, 2003, 2005; Saha and Shikin, 2004, 2005) 
in different context. Inhomogeneous cosmological models play an important role in 
understanding some essential features of the universe such as the formation of 
galaxies during the early stages of evolution and process of homogenization. The early 
attempts at the construction of such models have done by Tolman, (1934); and Bondi 
(1947) who considered spherically symmetric models. Inhomogeneous plane-symmetric 
model was first considered by Taub (1951, 1956) and later by Tomimura, (1978); 
Szekeres, (1975); Collins and Szafron, (1979a, 1979b);  Szafron and Collins, (1979). 
Bali and Tyagi, (1990) obtained a plane-symmetric inhomogeneous cosmological models 
of perfect fluid distribution with electro-magnetic field. \\

In this paper we study plane-symmetric inhomogeneous models in presence of perfect 
fluid with shear and vanishing acceleration. Among the cases with shear we find three 
classes of solutions.     
\section {Field Equations}
We consider the plane-symmetric line element in the general form
\begin{equation}
\label{eq1}
ds^{2} = - A^{2}(r,t)dt^{2} + B^{2}(r,t)dr^{2} + C^{2}(r,t)(dx^{2} + dy^{2}).
\end{equation}
The four velocity of the fluid has the form
\begin{equation}
\label{eq2}
u^{i} = \left(0, 0, 0, \frac{1}{A}\right).
\end{equation}
The energy-momentum tensor in the presence of a perfect fluid has the form
\begin{equation}
\label{eq3}
T_{ab} = (\rho + {p}) u_a u_b + {p} g_{ab},
\end{equation}
where $\rho$ and $p$ are the energy density and the pressure of the fluid, 
respectively. In this coordinate system the Einstein's field equation
\begin{equation}
\label{eq4}
G_{ab} \equiv R_{ab} - \frac{1}{2} g_{ab} R = - \kappa T_{ab}
\end{equation}
with (\ref{eq3}) read as 
\begin{equation}
\label{eq5}
G_{00} \equiv  \kappa \rho A^{2} = -\left(\frac{A}{B}\right)^{2} 
\left[\frac{2{C}''}{C} + \left(\frac{{C}'}{C}\right)^{2} - \frac{2{B}'{C}'}
{BC}\right] + \left(\frac{\dot{C}}{C}\right)^{2} + \frac{2\dot{B}\dot{C}}{BC},
\end{equation}
\begin{equation}
\label{eq6}
G_{11} \equiv  \kappa p B^{2} = \left(\frac{C'}{C}\right)^{2} + \frac{2A' C'}{A C} 
- \left(\frac{B}{A}\right)^{2}\left[\frac{2\ddot{C}}{C} + \left(\frac{\dot{C}}
{C}\right)^{2} - \frac{2\dot{A}\dot{C}}{AC}\right],
\end{equation}
\[
G_{22} \equiv  \kappa p C^{2} = \left(\frac{C}{B}\right)^{2}\left[\frac{A''}{A} + 
\frac{C''}{C} + \frac{A' C'}{A C} - \frac{{B}'{C}'}{BC} - \frac{{A}'{B}'}{AB}\right]
\]
\begin{equation}
\label{eq7}
-\left(\frac{C}{A}\right)^{2}\left[\frac{\ddot{B}}{B} + \frac{\ddot{C}}{C} + 
\frac{\dot{B}\dot{C}}{BC} - \frac{\dot{A}\dot{C}}{AC} - \frac{\dot{A}\dot{B}}{AB}\right]
\end{equation}
\begin{equation}
\label{eq8}
G_{01} \equiv 2\left[\frac{\dot{C}'}{C} - \frac{\dot{B}C'}{BC} - \frac{A' \dot{C}}
{AC}\right] = 0. 
\end{equation}
The dot denotes partial derivative with respect to time, the prime indicates partial 
derivative with respect to the coordinates $r$. $\kappa$ is the gravitational constant.
The consequences of the energy momentum conservation 
\begin{equation}
\label{eq9}
T^{ab}_{; b} = 0
\end{equation}
are the relations
\begin{equation}
\label{eq10}
p' = -(\rho + p)\frac{A'}{A}, \, \, \dot{\rho} = -(\rho + p)\left(\frac{\dot{B}}{B} 
+ \frac{2\dot{C}}{C}\right). 
\end{equation}
The plane symmetric solutions can be classified according to their four kinematical 
properties, i.e. rotation, acceleration, expansion and shear. In the comoving frame 
of reference these quantities read 
\begin{equation}
\label{eq11}
\omega_{ab} = u_{[a;b]} + \dot{u}_{[a^{u_{b}}]}\equiv 0, 
\end{equation}
\begin{equation}
\label{eq12}
\dot{u}_{i} = u_{i; n} u^{n} = \left(0, 0, \frac{A'}{A}, 0\right),
\end{equation}
\begin{equation}
\label{eq13}
\theta = u^{i}_{;i} = \frac{1}{A}\left(\frac{\dot{B}}{B} + \frac{2\dot{C}}{C}\right),
\end{equation}
\begin{equation}
\label{eq14}
\sigma_{i n} = u_(i; n) + \dot{u}_(i{u}_{n}) - \frac{1}{3}\theta (g_{in} + u_{i} u_{n}), 
\end{equation}
\begin{equation}
\label{eq15}
\sigma^{1}_{1} = \sigma^{2}_{2} = - \frac{1}{2}\sigma^{3}_{3} = \frac{1}{3A}
\left(\frac{\dot{C}}{C} - \frac{\dot{B}}{B}\right).
\end{equation}
The square bracket in Eq. (\ref{eq11}) denote antisymmetrization and the round 
bracket in Eq. (\ref{eq14}) denote symmetrization. 
\section{Plane symmetric Non-static solutions with shear and without acceleration}
To simplify the field equations we assume vanishing acceleration of the four velocity. 
According to Eqs. (\ref{eq10}) and (\ref{eq12}) the pressure of the fluid and the metric 
function $\ln(A)$ depend on time only, i. e.
\begin{equation}
\label{eq16}
p = p(t), \, \, A= A(t) 
\end{equation}
Eq. (\ref{eq9}) reads now
\begin{equation}
\label{eq17}
\dot{C}' = \frac{\dot{B} C'}{B}
\end{equation}
We have to investigate two essential different cases namely
\begin{equation} 
\label{eq18}
C' = 0,
\end{equation}
and the general case
\begin{equation}
\label{eq19}
C = C(r,t).
\end{equation} 
First of all we investigate the case $C = C(t)$.
\subsection{Solution for the case C = C(t), B = B(t)}
Without loss of generality, we can take $c = t$. The field equations take the form
\begin{equation}
\label{eq20}
\kappa \rho = \frac{1}{A^{2}}\left[\frac{1}{t^{2}} + \frac{2\dot{B}}{B t}\right],
\end{equation}
\begin{equation}
\label{eq21}
\kappa p = - \frac{1}{A^{2}}\left[\frac{1}{t^{2}} - \frac{2\dot{A}}{A t}\right],
\end{equation}
\begin{equation}
\label{eq22}
\kappa p = - \frac{1}{A^{2}}\left[\frac{\ddot{B}}{B} + \frac{\dot{B}}{B t} - 
\frac{\dot{A}}{A t} - \frac{\dot{A}\dot{B}}{AB}\right].
\end{equation}
Eq. (\ref{eq20}) defines the energy density. Eqs. (\ref{eq21}) and (\ref{eq22}) 
define the pressure. The only equation which actually has to be solved is the 
condition of consistency of Eqs. (\ref{eq21}) and (\ref{eq22})  
\begin{equation}
\label{eq23}
f(t) \dot{v} = g(t) v + 1.
\end{equation}
Here we have used the following abbreviations: 
\begin{equation}
\label{eq24}
v = \frac{1}{A^{2}},
\end{equation}
\begin{equation}
\label{eq25}
g(t) = 1 - \frac{\dot{B} t}{B} - \frac{\ddot{B} t^{2} }{B},
\end{equation}
\begin{equation}
\label{eq26}
f(t) = \frac{\dot{B} t^{2}}{2B} - \frac{1}{2}t.
\end{equation}
Eq. (\ref{eq23}) contains the two unknown quantities $A= A(t)$ and $B = B(t)$. 
We get all solutions of this class by choosing the function $B = B(t)$ arbitrarily.
Then we determine the function $v = \frac{1}{A^{2}}$ by solving the linear first 
order Eq. (\ref{eq23}). The solution reads  
\[
v = \frac{1}{A^{2}}
\]
\begin{equation}
\label{eq27}
= \exp{\left\{\int^{t}{\frac{g(t')}{f(t')}} dt' \right\}} \left[\int^{t}{
f^{-1}(t') \exp{\left \{- \int^{t'}{\frac{g(t'')}{f(t'')} dt''} \right\}}} dt' \right]
\end{equation}
The energy density, the pressure, and the shear can then be written as 
\begin{equation}
\label{eq28}
\rho = \frac{1}{\kappa t^{2}}\left[1 + \frac{2\dot{B} t}{B}\right] v,
\end{equation}
\begin{equation}
\label{eq29}
p = - \frac{1}{\kappa t^{2}}\left[1 - \frac{2\dot{A} t}{A}\right] v,
\end{equation}
\begin{equation}
\label{eq30}
\sigma^{1}~_{1} = \sigma^{2}~_{2} = - \frac{1}{2}\sigma^{3}~_{3} = \frac{1}{3 A t}
\left(1 - \frac{\dot{B} t}
{B}\right).
\end{equation}
Out of the infinite number of exact solutions of Kantowski-Sachs class we present only 
the solution following Mc Vittie and Wiltshire (1975). Let us choose
\begin{equation}
\label{eq31}
B = n \ln {t},
\end{equation}
where $n$ is constant, $n^{2} \ne 1$. The functions $f(t)$, $g(t)$ and $v(t)$ read
\begin{equation}
\label{eq32}
f(t) = \frac{1}{2} t (n - 1), 
\end{equation}
\begin{equation}
\label{eq33}
g(t) = 1 - n^{2},
\end{equation}
\begin{equation}
\label{eq34}
v(t) = \frac{1}{A^{2}} = \frac{1}{n^{2} - 1} + C_{0}t^{-2(n + 1)},
\end{equation}
where $C_{0}$ is an arbitrary constant. In this case the geometry of the universe 
(\ref{eq1}) takes the form as
\begin{equation}
\label{eq35}
ds^{2} = - \left[\frac{(n^{2} - 1)}{1 + C_{0}(n^{2} - 1)t^{-2(n + 1)}}\right] dt^{2} 
+ (n \ln{t})^{2} dr^{2} + t^{2}(dx^{2} + dy^{2}).
\end{equation}
The energy density, the pressure and the shear for the model (\ref{eq35}) are given 
by
\begin{equation}
\label{eq36}
\rho = \frac{1}{\kappa t^{2}}\left(1 + \frac{2}{\ln {t}}\right) \left[\frac{1}{(n^{2} - 1)} 
+ \frac{C_{0}}{t^{2(n + 1)}}\right],
\end{equation}
\begin{equation}
\label{eq37}
p = \frac{(4n + 3)}{\kappa t^{2}}\left[\frac{1}{(n^{2} - 1)} + \frac{C_{0}}
{t^{2(n + 1)}}\right], 
\end{equation}
\begin{equation}
\label{eq38}
\sigma^{1}~_{1} = \sigma^{2}~_{2} = - \frac{1}{2}\sigma^{3}~_{3} = \frac{1}{3t} 
\left(1 - \frac{1}{\ln {t}}\right) \left[\frac{1}{(n^{2} - 1)} 
+ \frac{C_{0}}{t^{2(n + 1)}}\right]. 
\end{equation}
From Eqs. (\ref{eq36}) and (\ref{eq37}), we observe that $\rho > 0$, $p > 0$ and 
$\rho$ is a decreasing function of time. This model satisfies the weak and strong 
energy conditions and also has a physically acceptable fall-off behaviour in both 
$r$ and $t$.
\subsection{Solution for the case C = C(t), B = B(r,t)}
In this case we write down the Einstein's field equations as follows:
\begin{equation}
\label{eq39}
\kappa \rho = \frac{1}{A^{2} \, t^{2}}\left[1 + \frac{2 \dot{B} \, t}{B}\right], 
\end{equation}
\begin{equation}
\label{eq40}
\kappa p = - \frac{1}{A^{2}}\left[\frac{1}{t^{2}} - \frac{2\dot{A}}{A \, t}\right],
\end{equation}
\begin{equation}
\label{eq41}
\kappa p = -\frac{1}{A^{2}}\left[\frac{\ddot{B}}{B} + \frac{\dot{B}}{B \, t} - \frac{\dot{A}}
{A \, t} - \frac{\dot{A} \dot{B}}{A B}\right].
\end{equation}
From Eqs.(\ref{eq40}) and (\ref{eq41}), by using the condition of consistency, we have 
\begin{equation}
\label{eq42}
\ddot{B} + \dot{B}\left(\frac{1}{t} - \frac{\dot{A}}{A}\right) + \left(\frac{\dot{A}}{A \, t} 
- \frac{1}{t^{2}}\right)B = 0.
\end{equation}
Equation (\ref{eq42}) is a non-linear partial differential equation. To solve this equation, 
following the technique of Herlt (1996), we choose
\begin{equation}
\label{eq43}
\frac{1}{A^{2}} = \frac{1}{n^{2} - 1} + C_{0} t^{-2(n + 1)}, \, \, n^{2} \ne 1
\end{equation}
where $C_{0}$ is an integrating constant. Using Eq. (\ref{eq43}) in (\ref{eq42}), we obtain 
\begin{equation}
\label{eq44}
\ddot{B} + P \dot{B} - N B = 0,
\end{equation}
where
\begin{equation}
\label{eq45}
P = \frac{t^{-1} - n(n^{2} - 1) C_{0}t^{-(2n + 3)}}{1 + C_{0}(n^{2} - 1) t^{-2(n + 1)}},
\end{equation}
\begin{equation}
\label{eq46}
N = \frac{P}{t} = \frac{n^{2} t^{-2} - n(n^{2} - 1) C_{0}t^{-2(n + 2)}}{1 + C_{0}
(n^{2} - 1) t^{-2(n + 1)}}.
\end{equation}
Eq. (\ref{eq44}) has the special solution $B = t^{n}$. Following D'Alembert's method. We 
insert the expression 
\begin{equation}
\label{eq47}
B = D(r,t) t^{n}
\end{equation}
in to Eq. (\ref{eq44}), where $D(r,t)$ is a function of $r$ and $t$, we obtain
\begin{equation}
\label{eq48}
\ddot{D} + (P + 2n t^{-1}) \dot{D} + (n - 1) P (1 + n t^{-1})D = 0
\end{equation}
Integrating Eq. (\ref{eq48}), we obtain
\[
D = C_{1} e^{-\frac{P t}{2}} t^{-n}\, Whittaker \, M \left[\frac{-I\sqrt{P} n(n - 2)}
{\sqrt{4n - 4 - P}}, n - \frac{1}{2}, I\sqrt{P} \sqrt{4n - 4 -P} t \right]
\]
\begin{equation}
\label{eq49}
+ \, C_{2} e^{-\frac{P t}{2}} t^{-n}\, Whittaker \,  W \left[\frac{-I\sqrt{P} n(n - 2)}
{\sqrt{4n - 4 - P}}, n - \frac{1}{2}, I\sqrt{P} \sqrt{4n - 4 -P} t \right].
\end{equation}
From Eqs. (\ref{eq47}) and (\ref{eq49}), we have
\[
B = C_{1} e^{-\frac{P t}{2}}\, Whittaker \, M \left[\frac{-I\sqrt{P} n(n - 2)}
{\sqrt{4n - 4 - P}}, n - \frac{1}{2}, I\sqrt{P} \sqrt{4n - 4 -P} t \right]
\]
\begin{equation}
\label{eq50}
+ \, C_{2} e^{-\frac{P t}{2}} \, Whittaker \,  W \left[\frac{-I\sqrt{P} n(n - 2)}
{\sqrt{4n - 4 - P}}, n - \frac{1}{2}, I\sqrt{P} \sqrt{4n - 4 -P} t \right].
\end{equation}
The energy density, the pressure, and the shear can be computed easily now by means of 
formulae (\ref{eq5}), (\ref{eq6}) and (\ref{eq15}). The geometry of the universe 
(\ref{eq1})in this case reduces to
\begin{equation}
\label{eq51}
ds^{2} = - A^{2} dt^{2} + B^{2} dr^{2} + t^{2} (dx^{2} + dy^{2}),
\end{equation}
where $A$ and $B$ are given by (\ref{eq43}) and (\ref{eq50})
\subsection{Solution for the general case C = C(r,t), B = B(r,t)}
The pressure $p$ and the function $A$ depend on the time $t$ only. By means of a 
scaling of the coordinate $t$ we arrive at
\begin{equation}
\label{eq52}
A = 1, \, \, \,  \dot{C}'= \frac{\dot{B}C'}{B}.
\end{equation}
In this case we write down the field equations as follows: 
\begin{equation}
\label{eq53}
\kappa \rho = - \frac{1}{B^{2}}\left[\frac{2C''}{C} + \frac{C'^{2}}{C^{2}} - \frac{2B' C'}
{B C}\right] + \frac{\dot{C}^{2}}{C^{2}} + \frac{2\dot{B}\dot{C}}{BC},
\end{equation}
\begin{equation}
\label{eq54}
\kappa p = \frac{C'^{2}}{C^{2}B^{2}} - \left[\frac{2\ddot{C}}{C} + \frac{\dot{C^{2}}}
{C^{2}}\right]
\end{equation}
\begin{equation}
\label{eq55}
\dot{C}'= \frac{\dot{B}C'}{B}.
\end{equation}
The second equation for the pressure follows after differentiation of Eq. (\ref{eq54}). 
We can integrate Eq. (\ref{eq55}) to obtain
\begin{equation}
\label{eq56}  
\ln {C'} = \ln {B} + \frac{1}{2} \ln {[1 - \epsilon \hbar^{2}(r)]}.
\end{equation}
Here $\frac{1}{2} \ln{[1 - \epsilon \hbar^{2}(r)]}$ is the integration ``constant'', 
where $\hbar(r)$ is an arbitrary function of $r$. We get
\begin{equation}
\label{eq57}
\frac{{C'}^{2}}{B^{2}} = 1 - \epsilon \hbar^{2}(r).  
\end{equation}
The remaining equation to be solved is (\ref{eq54}). We use the result (\ref{eq57}) 
and get
\begin{equation}
\label{eq58}
p(t) = \frac{1}{\kappa C^{2}}\left[1 - \epsilon \hbar^{2}(r) - \{2C\ddot{C} + 
\dot{C}^{2}\}\right],
\end{equation}
with the abbreviation 
\begin{equation}
\label{eq59}
C(r,t) = Z^{\frac{2}{3}}(r,t),
\end{equation}
we obtain from Eq. (\ref{eq58}) the nonlinear differential equation
\begin{equation}
\label{eq60}
\ddot{Z} + \frac{3}{4}\kappa p Z = \frac{3}{4}\left(1 - \epsilon \hbar^{2}(r)\right) 
Z^{-\frac{1}{3}}.
\end{equation}
The mass density follows from eq. (\ref{eq53}) by the use of (\ref{eq57}) and (\ref{eq58}) 
\[
\kappa  \rho = \frac{2}{C}\left[\left(1 - \epsilon \hbar^{2}(r)\right)C'' + 
\hbar(r)\hbar'(r) C'\right] - \kappa p (2C'' C + C'^{2}) +
\]
\begin{equation}
\label{eq61}
\dot{C}^{2}\left(\frac{2C''}{C} + \frac{C'^{2}}{C^{2}} + \frac{1}{C^{2}}\right) + 
\frac{2\dot{C}'\dot{C}}{C'C} + \ddot{C} \left(4C'' + \frac{2C'^{2}}{C}\right)
\end{equation}
To obtain solutions we have to proceed as follows. First of all we choose some $p(t)$ 
{\it ad hoc}. Then we have to solve Eq. (\ref{eq60}). The last step is to the calculation of 
mass density and shear. Eq. (\ref{eq60}) can be simplified by means of the abbreviations  
\begin{equation}
\label{eq62}
Z(r,t) = E \hat{Z}(r,t), \, \, E^{\frac{4}{3}} = \frac{3}{4} \left(-\frac{1}{\epsilon} 
+ \hbar^{2}(r)\right).
\end{equation}
Then we get
\begin{equation}
\label{eq63}
\ddot{\hat{Z}} + \frac{3}{4}\kappa p \hat{Z} = -\epsilon \hat{Z}^{-\frac{1}{3}}
\end{equation}
with
\begin{equation}
\label{eq64}
C(r,t) = \sqrt{ \frac{3}{4} \left(-\frac{1}{\epsilon} + \hbar^{2}(r)\right)}\, 
\hat{Z}^{\frac{2}{3}}.
\end{equation}
The regularity conditions at the base of the plane are fulfilled automatically 
if the arbitrary function $\hbar(r)$ is chosen as 
\begin{equation}
\label{eq65}
\hbar(r) = 0 \, \, \, at \, \, r = 0.
\end{equation}
In this case Eq. (\ref{eq61}) reduces to
\[
\kappa  \rho = \frac{2C''}{C} - \kappa p (2C'' C + C'^{2}) + \dot{C}^{2}\left(\frac{2C''}
{C} + \frac{C'^{2}}{C^{2}} + \frac{1}{C^{2}}\right) + 
\]
\begin{equation}
\label{eq66}
\frac{2\dot{C}'\dot{C}}{C'C} + \ddot{C} \left(4C'' + \frac{2C'^{2}}{C}\right)
\end{equation}
At first glance the solution of (\ref{eq63}) seems rather trivial. We can choose an 
arbitrary function $\hat{Z} = \hat{Z}(t)$ and then determine the pressure. But this 
does not work. First of all notice that in the case $\hat{Z} = \hat{Z}(t) \alpha(r)$ 
the function $C(r,t)$ has the form
\begin{equation}
\label{eq67}
C(r,t) = \sqrt{\frac{3}{4} \left(-\frac{1}{\epsilon} + \hbar^{2}(r)\right)} \, \alpha^{\frac{2}
{3}}(r) \hat{Z}^{\frac{2}{3}}(t).
\end{equation}
As a result, the shear vanishes because 
\begin{equation}
\label{eq68}
\sigma^{1}\, \, _{1} = \sigma^{2}\, \, _{2} = - \frac{1}{3} \sigma^{3}\, \, _{3} = 
\frac{1}{3}\left(\frac{\dot{C}}{C} - \frac{\dot{B}}{B}\right) =  - \frac{1}{3}\frac{d}{dt}
\left(\ln{\left[\frac{C'}{C}\right]}\right). 
\end{equation}
Therefore $\hat{Z}$ has to depend on the co-ordinate $r$ in the form
\begin{equation}
\label{eq69}
\hat{Z} = \hat{Z}\left\{J_{1}(r), J_{2}(r), t \right\}   \, \, for \, \,  \epsilon \ne 0.
\end{equation}
Here the arbitrary functions $J_{1}(r)$, and $J_{2}(r)$ are integration ``constant'' which 
enter the general solution of Eq. (\ref{eq63}). We see that the r-dependence in the solution 
comes from these integration ``constant''. 

If we set $\epsilon = -1$ and $\hbar(r) = 0$, then by (\ref{eq56}) 
\begin{equation}
\label{eq70}
C' = B.
\end{equation}
Hence, in this case, the metric (\ref{eq1}) reduces to  
\begin{equation}
\label{eq71}
ds^{2} = - dt^{2} + {C'}^{2}(r,t)dr^{2} + C^{2}(r,t)(dx^{2} + dy^{2}).
\end{equation} 
For model (\ref{eq71}), the corresponding expressions for pressure and density can be computed 
from Eqs. (\ref{eq63}) and (\ref{eq66}) respectively. \\

Let us recall here two important families of solutions corresponding to particular 
choices of $p(t)$ (we do not claim to be exhaustive). 

\subsubsection{The case $p$ = constant $\ne 0$}
In this case the differential equation (\ref{eq63}) reduces to the form
\begin{equation}
\label{eq72}
\ddot{\hat{Z}} + N \hat{Z} = -\epsilon \hat{Z}^{-\frac{1}{3}}
\end{equation} 
where $N = \frac{3}{4}\kappa p$. Integrating (\ref{eq72}), we get expression for $\hat{Z}$
\begin{equation}
\label{eq73}
\frac{N}{6} \hat{Z}^{3} + \frac{9}{10} \epsilon \hat{Z}^{\frac{5}{3}} + (1 - k_{1}) 
\hat{Z} - k_{2} = 0,
\end{equation}
where $k_{1}$ and $k_{2}$ are integrating constant. Putting the value of $\hat{Z}$ 
from (\ref{eq73}) in (\ref{eq67}), we get the value of $C(r,t)$ which is not reported 
here due to complexity.
\subsubsection{Dust case $p$ = 0}
In this case the differential equation (\ref{eq63}) reduces to the form
\begin{equation}
\label{eq74}
\ddot{\hat{Z}} = -\epsilon \hat{Z}^{-\frac{1}{3}}.
\end{equation} 
Integrating (\ref{eq74}), we get expression for $\hat{Z}$
\begin{equation}
\label{eq75}
(1 - m_{1}) \hat{Z} + \frac{9}{10} \epsilon \hat{Z}^{\frac{5}{3}} - m_{2} = 0 
\end{equation}
where $m_{1}$ and $m_{2}$ are integrating constants. Putting the value of $\hat{Z}$ 
from (\ref{eq75}) in (\ref{eq67}), we get the value of $C(r,t)$.

\subsubsection{The Explicit Form of the General Solution}
As stated in the preceding section, a solution of field equations (\ref{eq53})-(\ref{eq55})
 can be obtained for any given form of $p(t)$. We give here the explicit form of the 
general solution with no restriction on $p(t)$. \\ 
 
Let us perform in (\ref{eq63}) the following substitution
\begin{equation}
\label{eq76}
\epsilon = 0
\end{equation}
so that we obtain the solution
\begin{equation}
\label{eq77}
\ddot{\hat{Z}} + \frac{3}{4}\kappa p \hat{Z} = 0,
\end{equation}
which is linear in $\hat{Z}$. The general equation of (\ref{eq77}) is of the form 
\begin{equation}
\label{eq78}
\hat{Z} =  D_{1} a(t) + D_{2} b(t) \, \, for \, \, \epsilon = 0,
\end{equation}
where $D_{1}$ and $D_{2}$ are arbitrary functions of $r$ and $a$ and $b$ are two different 
(nontrivial) particular solutions of (\ref{eq77}). In fact, we need only know one nontrivial 
particular solution of (\ref{eq77}), namely $a(t)$. This is because both $a$ and $b$ verify 
(\ref{eq77}), that is 
\begin{equation}
\label{eq79}
\ddot{a} + \frac{3}{4} \kappa p(t) a = 0,
\end{equation} 
and the analogous expression for b(t). Both functions must then be related to one and 
another by
\begin{equation}
\label{eq80}
\frac{\ddot{a}}{a} = \frac{\ddot{b}}{b},
\end{equation}
which can be solved, up to a quadrature, for $b$ once $a(t)$ is known. A particular choice 
of $b(t)$, different from $a(t)$, is given by 
\begin{equation}
\label{eq81}
b(t) = a(t) \int^{t}{a^{-2}} dt'
\end{equation}
Note that (\ref{eq79}) actually may be interpreted as a mere definition of $p(t)$ 
(because $a$ depends only on time). We can consider either $p(t)$  being arbitrary 
and $a(t)$ derived through (\ref{eq79}), or vice verse. 
\section{Conclusion}
We have studied a new class of plane symmetric inhomogeneous cosmological models 
with shear and vanishing acceleration in presence of a perfect fluid. Our analysis 
is an attempt to obtain more exact solutions so that our understanding of these 
objects may be improved. It is hoped that some of the solutions presented here will 
provide the basis for a detailed physical analysis of plane symmetric inhomogeneous 
models with shear and vanishing acceleration in the gravitational context. The 
solutions we had obtained for the case $p \ne 0$ and $p = 0$ (dust case) and explore 
the consistency of the formalism and outlined obtaining generalization. Some of the 
models are being persued at present and would be reported in future work.  
\section*{Acknowledgements}
Authors (A. Pradhan and P. Pandey) would like to thank the Inter-University Centre for 
Astronomy and Astrophysics, Pune, India for providing facility where this paper was 
carried out.  

\end{document}